\documentclass[fleqn,11pt]{manuscript}

\usepackage{soul}
\usepackage{relsize}

\newcommand{\Fig}{Fig.}
\newcommand{\Figs}{Figs.}
\newcommand{\Eq}{Eq.}
\newcommand{\Eqs}{Eqs.}
\newcommand{\Sec}{Sec.}
\newcommand{\Secs}{Secs.}

\newcommand{\SW}{SW}
\newcommand{\SF}{SF}
\newcommand{\SWSF}{SW-SF}
\newcommand{\SFSW}{SF-SW}

\newcommand{\physical}{face-to-face}

\newcommand{\Pinf}{\beta} 
\newcommand{\Prec}{\gamma} 
\newcommand{\thres}{\Theta} 
\newcommand{\comb}{\alpha} 
\newcommand{\npi}{\eta} 
\newcommand{\grf}{\delta}  
\newcommand{\trust}{\omega} 

\newcommand{\sfexp}{\nu} 

\newcommand{\degOpi}{{k_i}} 
\newcommand{\degEpi}{{l_i}} 
\newcommand{\nbsI}{l_i^I} 
\newcommand{\nbsA}{k_i^A} 

\newcommand{\Pit}[2]{P_i^{#1\rightarrow~#2}}
\renewcommand{\t}{[t]}
\newcommand{\tm}{[t-1]}
\newcommand{\Ca}{C_\alpha(i)}
\newcommand{\Cua}{C^U_\alpha(i)}
\newcommand{\Caa}{C^A_\alpha(i)}

\graphicspath{./figures/output/}

\newcommand{\heatmapsubfig}[1]{%
	\begin{subfigure}{0.47\textwidth}
		\centering
		\vspace{-3pt}
		\includegraphics[width=0.85\textwidth]{figures/output/#1-0}
		\vspace{-6pt}
		\caption{}%
		\label{subfig:#1}%
	\end{subfigure}
}

\newcommand{\heatmaprow}[2]{%
	\hspace*{\fill}
	\heatmapsubfig{#1}
	\hspace*{\fill}
	\heatmapsubfig{#2}
	\hspace*{\fill}
}

\newcommand{\heatmapsquare}[4]{%
	\heatmaprow{#1}{#2} \\
	\heatmaprow{#3}{#4}
}

\newcommand{\graphsubfig}[1]{%
	\begin{subfigure}{0.425\textwidth}
		\centering
		\caption{}%
		\label{subfig:#1}
		\includegraphics[height=140pt]{figures/output/#1-0}
	\end{subfigure}
}

\newcommand{\graphrow}[2]{%
	\hspace{-20pt}
	\graphsubfig{#1}
	\hspace{20pt}
	\graphsubfig{#2}
}

\newcommand{\graphsquare}[4]{%
	\graphrow{#1}{#2} \\
	\graphrow{#3}{#4}
}

\title{Epidemic risk perception and social interactions lead to awareness cascades on multiplex networks}

\author[1\orcidlink{0000-0002-2105-8805}]{Tim Van Wesemael}
\author[2,3\orcidlink{0000-0001-9046-8739}]{Luis E.C. Rocha}
\author[1\orcidlink{0000-0003-4084-9992}]{Jan M. Baetens}
\affil[1]{BionamiX, Department of Data Analysis and Mathematical Modelling, Ghent University}
\affil[2]{Department of Economics, Ghent University}
\affil[3]{Department of Physics and Astronomy, Ghent University}
\corrauthor[1]{Tim Van Wesemael}{tim.vanwesemael@ugent.be}


\begin{abstract}
	The course of an epidemic is not only shaped by infection transmission over \physical~contacts, but also by preventive behaviour caused by risk perception and social interactions.
	This study explores the dynamics of coupled awareness and biological infection spread within a two-layer multiplex network framework.
	One layer embodies \physical~contacts, with a biological infection transmission following a simple contagion model, the SIR process.
	Awareness, modelled by the linear threshold model, a complex contagion, spreads over a social layer and induces behaviour that lowers the chance of a biological infection occurring.
	It may be provoked by the presence of either aware or infectious neighbours.
	We introduce a novel model combining these influences through a convex combination, creating a continuum between pure social contagion and local risk perception.
	Simulation of the model shows distinct effects arising from the awareness sources.
	Also, for convex combinations where both input sources are of importance, awareness cascades that are not attributable to only one of these sources, emerge.
	Here, the combination of a small-world \physical~and a scale-free social layer, but not vice versa, cause the number of infections to decrease with increasing transmission probability.
\end{abstract}

\begin{document}
	\flushbottom
	\maketitle
	\thispagestyle{empty}

	\section{Introduction}%
	\label{sec:introduction}

	Some infectious diseases spread through networks of \physical~contacts~\cite{PastorSatorras2015}.
	To reduce the spread of such diseases, preventive measures such as hand washing~\cite{Andrews2015}, or pharmaceutical interventions such as vaccination~\cite{Wang2017a}, are efficient methods.
	However, the decision to follow specific recommendations from authorities happens at individual level and might be affected by mass media, information circulating through social networks, and the individual's own perception of the epidemic~\cite{Funk2010, Weston2018, Kobayashi2022a, Epstein2022, Silva2023}.
	An individual may become aware of the disease either because of the sickness of its face-to-face contacts~\cite{Bagnoli2007}, or because it is informed through its social interactions~\cite{Kabir2020}.
	In the latter case, the awareness spreads through its social network, similar to biological infections that spread through the \physical~one~\cite{Granell2014}.
	For example, it has been argued that the rapid dissemination of information on SARS in 2003 encouraged preventive behaviour and helped to contain the epidemic~\cite{Durham2011, Shi2019}, and during the recent COVID-19 pandemic, press coverage and social media attention to the virus were so prevalent that the term ``infodemic'' was introduced~\cite{Medford2020}.
	We devise and study an epidemic model on social networks, where not only the transmission of an infectious disease is considered, but also the effect of the concurrent diffusion of awareness of the disease.

	An infectious disease may spread among spatially close contacts according to SIR dynamics~\cite{Funk2009}.
	This is a simple contagion; i.e.\ one described by pairwise interactions, which implies that the probability that a vertex infects a neighbour does not depend on the state of any other neighbour\ \cite{Horsevad2022}.
	Additionally, some individuals may take precautions against infection, either because they notice that their \physical~contacts are sick, or because they hear of the disease from their social contacts.
	In contrast to the biological infections, this process can be modelled by a complex contagion~\cite{Guilbeault2018}.
	Here, the awareness state of a vertex is not solely determined by a pairwise interaction, but rather by the states of multiple vertices in the neighbourhood~\cite{Campbell2013,Centola2018}.
	The linear threshold model can represent this type of interaction~\cite{Watts2002,Guo2015}.
	In this model, each vertex has an associated threshold and becomes aware (more generally, active) if the aware proportion of its neighbourhood exceeds the threshold, here denoted as \( \thres \).
	Originally proposed as a behavioural model~\cite{Granovetter1978}, the threshold model gained traction for its ability to activate the (almost) complete network from a small seed~\cite{Watts2002, Gleeson2007}.
	Such an event is referred to as a global cascade, so in the context of our study, we speak of an awareness cascade when (almost) the whole population is aware in steady-state.
	In addition to the final outcome, the transient course of the number of aware vertices is of importance in this study, because we consider its co-evolution with the spread of an infectious disease.
	This process has been studied previously under the name of local awareness contagion spreading\ \cite{Guo2015, Wang2019a}.
	Next to this model, the co-evolution between biological and social contagions --- on and off networks --- has been studied under many other assumptions~\cite{Verelst2016, Wang2019}.

	When using a threshold model in a coupled infection-awareness setting, two quantities can serve as input: the infectious proportion and the aware proportion of neighbours.
	The former case is referred to as local risk perception~\cite{Bagnoli2007}, prevalence-based perception~\cite{Verelst2016}, rational response~\cite{Fu2017}, or self-initiated awareness~\cite{Kan2017}; the latter case as belief-based perception~\cite{Verelst2016}, social contagion~\cite{Fu2017} or awareness diffusion~\cite{Kan2017}.
	A combination of these two input sources has been taken into account when modelling behavioural responses to epidemics.
	For example, Mao et al.~\cite{Mao2011, Mao2011a, Mao2012} use a threshold model for preventive behaviour adoption considering both the disease and awareness state of neighbours.
	They work with two separate thresholds: one for the infection ratio, and one for the proportion of neighbours that have taken prophylactic measures.
	Only one has to be exceeded for the agent to take preventive steps.
	Also, for a continuous awareness state a variant of the convex combination of the social and \physical~neighbourhood states, as used here, albeit for a discrete state, has been described earlier.
	Awareness diffuses according to the dynamics proposed by Abelson~\cite{Fu2017} or DeGroot~\cite{Paarporn2016, Paarporn2017}.
	These are however not complex contagions, as we use.
	The model of Du et al.~\cite{Du2021} builds on this.
	It includes a threshold for the continuous awareness, above which protective behaviour is adopted.
	Finally, a similar approach is one in which the proportion in one layer plays a role, and that layer is decided with a probability at each time step for each vertex independently\ \cite{Wang2021b}.

	In this study, a two-layer multiplex network models the population structure on which the coupled infection-awareness process takes place~\cite{Granell2013,Guo2015}.
	A member of the population is represented by two vertices in a discrete state, one on each layer.
	In the \physical~layer, the edges portray interactions with physical proximity and the states express the infection status.
	On this layer, an infection that spreads to physically close vertices through the SIR model takes place.
	Each susceptible vertex (S) might become infectious (I) because of a neighbour in that state.
	Infectious vertices heal and become recovered (R).
	It is assumed that once in this state, they are immune and cannot become infectious again.
	A seasonal influenza, for example, can be modelled by this process~\cite{Fu2017}.

	In the social layer, the edges of the network represent social connections and the state of a vertex represents its awareness.
	An aware vertex (A) takes precautions, reducing the probability of becoming infectious, whereas an unaware one (U) does not.
	The awareness state of a vertex is determined by a threshold model, with threshold \( \thres \).
	Both the proportion of aware neighbours, and the proportion of infectious neighbours serve as input.
	The combination parameter \( \comb \) controls a convex combination between these sources.
	In an extension we include trusted institutions in the model, which are static vertices that are always aware or unaware.
	However there are no mandated behavioural changes, so the model does not represent highly-infectious diseases such as COVID-19.

	Our model shows a difference between pure social contagion and local risk perception.
	The effects of the former result in a situation where, in the steady state, either every vertex is aware, or none is.
	For the latter, there are intermediate effects that depend on the parameter combinations.
	In addition, when there is a mix of the two awareness sources, the model exhibits complex behaviour.
	It can produce awareness cascades not only from the initial condition, but also from the transient number of infectious vertices, which are able to transition their neighbours to the aware state.
	These, in turn, form an initial seed, from which awareness can spread over the entire network.
	Such seed is more likely to form in a scale-free network than in a small-world network, and thus the effects of awareness are more pronounced on the former than on the latter.
	The same holds for the effects of trusted awareness sources.
	The sources themselves may stop the infection from spreading when the social layer is scale-free, however a small-world social layer benefits from also considering the infectious neighbourhood to stop the infection.

	In \Sec~\ref{sec:networks} the single-layer and multiplex networks consisting of Watts-Strogatz small-world networks and scale-free configuration models are introduced.
	Afterwards, in \Sec~\ref{sec:dynamics} the rules according to which the infections and awareness spread and interact are presented.
	In \Sec~\ref{sec:setup} the set-up of our computational experiments is explained, of which the results are discussed in \Sec~\ref{sec:results}.
	We conclude with our main findings in \Sec~\ref{sec:conclusion}.

	\section{Methods}%
	\label{sec:methods}

	\subsection{Networks}\label{sec:networks}

\begin{figure*}[t]
		\centering
		\hfill
		\begin{subfigure}{0.38\textwidth}
			\centering
			\includegraphics[width=\textwidth]{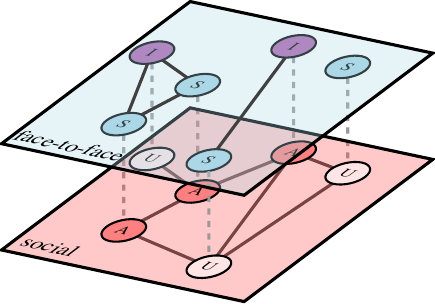}
			\caption{}%
			\label{subfig:network}
		\end{subfigure}
		\hfill
		\begin{subfigure}{0.25\textwidth}
			\centering
			\includegraphics[width=\textwidth]{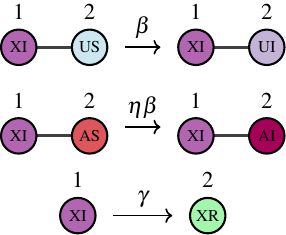}
			\caption{}%
			\label{subfig:epi_dynamics}
		\end{subfigure}
		\hspace*{\fill}
		\\
		\hfill
		\begin{subfigure}{0.23\textwidth}
			\centering
			\includegraphics[width=\textwidth]{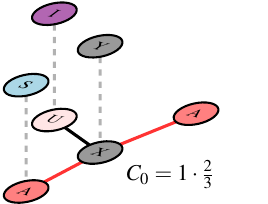}
			\caption{}%
			\label{subfig:opi_dynamics_0}
		\end{subfigure}
		\hfill
		\begin{subfigure}{0.23\textwidth}
			\centering
			\includegraphics[width=\textwidth]{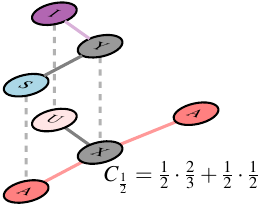}
			\caption{}%
			\label{subfig:opi_dynamics_5}
		\end{subfigure}
		\hfill
		\begin{subfigure}{0.23\textwidth}
			\centering
			\includegraphics[width=\textwidth]{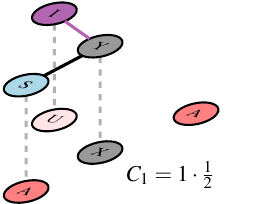}
			\caption{}%
			\label{subfig:opi_dynamics_10}
		\end{subfigure}
		\hfill
		\caption{%
			An illustration of the coupled infection-awareness model:
			(\subref{subfig:network}) a two layer network with a \physical~and social contact layers; dashed lines show that the same vertex is present in each layer, whereas full lines shows the network edges in the respective layers.
			(\subref{subfig:epi_dynamics}) The infection transitions: infection of an unaware neighbour, infection of an aware neighbour, and recovery.
			The convex combination given by \Eq~\eqref{eq:convex} for (\subref{subfig:opi_dynamics_0}) \( \comb=0 \),  (\subref{subfig:opi_dynamics_5}) \( \comb=0.5 \) and (\subref{subfig:opi_dynamics_10}) \( \comb=1 \).
			\( X \) and \( Y \) represent an arbitrary awareness, respectively infection state, \( X \in \{ U, A \}, Y \in \{ S, I, R \} \).
		}%
		\label{fig:model}
	\end{figure*}

	A network is defined by a set of \( N \) vertices that are connected by edges \( (i,j) \) with \( i \) and \( j \) being vertices.
	In an undirected network all interactions are symmetric, thus, if \( i \) can infect \( j \), then \( j \) can infect \( i \) as well.
	If an edge \( (i, j) \) exists, \( i \) is called a neighbour of \( j \) and vice versa.
	The number of neighbours of a vertex is referred to as its degree \( k \).
	Other network properties of interest are clustering, a measure of how many neighbours of a vertex are connected between themselves, and network diameter, the longest shortest path between any two vertices~\cite{Newman2018}.
	We adopt a multiplex network as underlying topology for our coupled infection-awareness model.
	This is a multilayer network in which a bijection exists between vertices of different layers~\cite{Boccaletti2014}.
	More specifically, two layers are used, one for the social and one for the \physical~connections.
	An edge can appear in none, one, or both of the layers (\Fig~\ref{subfig:network}).

	We will consider two network models in our study.
	One is a small-world model, the other one is a scale-free model.
	Both are relatively simple, and do not capture all the complexities of real social networks, but still have interesting properties~\cite{Newman2018,Broido2019,Ozella2021}.
	A small-world network exhibits high clustering, mirroring the tendency of individuals to have contacts in common with their other contacts.
	The scale-free model has a heterogeneous degree distribution, with a few vertices having a high degree, and many having a low degree, allowing the inclusion of a few well-connected individuals in the model.
	The goal here is to study how these two specific topologies, with their respective properties, regulate the dynamics.

	In the Watts-Strogatz small-world model (SW) model, vertices are initially positioned uniformly spaced in a circle and are connected to their \( K \) nearest neighbours, where \( K \) is a model parameter.
	This structure has a high level of clustering.
	In the second step, edges are chosen with probability \( p_r \), the second model parameter, and rewired by changing one of the vertices in the link randomly.
	This procedure creates long-distance connections and reduces the diameter of the network.
	Here we use \( K=6 \) and adopt a rewiring probability \( p_r \) of  \( 0.1 \).
	Note that since no edges are created or removed, the mean degree remains equal to \( K \).

	The second network model is the configuration model with a power-law degree distribution (SF), \( P(k) \propto k^{-\sfexp} \).
	We adapt the distribution such that each vertex has a degree of at least three~\cite{Clauset2009}, i.e.
	\begin{equation}
		P(k) \propto
		\begin{cases}
			0, & k < 3, \\
			k^{-\sfexp}, & k \geq 3.
		\end{cases}\label{eq:scalefree}
	\end{equation}
	Such a degree distribution leads to a few vertices that are highly connected, and many with low degree.
	The \SF{}~model assigns a degree to each vertex sampled from the given distribution and then randomly connects available vertices, until there are no possible edges left.
	This results in a network with the desired degree distribution, but with a random topology.
	In order to obtain networks with mean degree 6, like the SW network above, we set \( \sfexp=2.72 \).
	The heterogeneous degree distribution of the \SF{}~network and the high clustering of the \SW{}~one, influence the simple and complex contagion each in their own way, allowing for a revealing comparison\ \cite{Centola2018, Wang2019}.

	In our model, one layer represents the \physical~interactions and the second layer the social interactions between vertices.
	An individual is thus represented by two vertices, one in each layer.
	We study two configurations.
	In the first case, both layers have the same network structure, i.e.~the \physical~and social networks perfectly coincide.
	Hence, this is referred to as a single-layer network.
	In the second configuration, the network structure is generated independently on each layer, and thus the topologies are uncorrelated.
	We arrive at two multiplex network models.
	One where the \physical~layer is a \SF{}~network, and the social layer is a \SW{}~network, and another one where the network layers are mutually swapped.
	For the sake of brevity, we first refer to the \physical~layer and then to the social layer, so the former multiplex network is referred to as a \SFSW~network.

	\subsection{Dynamics}\label{sec:dynamics}
	Two dynamic, interacting processes take place on the networks.
	The first one represents the spread of an infection (on the face-to-face layer) and the second the spread of awareness (on the social layer).
	Each vertex in the network is associated with an infection or awareness state, depending on the layer it belongs to.
	In the biological infection model, a vertex can be in the susceptible, infectious or recovered state.
	At each time step \( t \), an infectious vertex recovers with probability \( \Prec \), and infect a susceptible neighbour with probability \( \Pinf \), resulting in SIR dynamics~\cite{Lu2023}.
	In the awareness model, the vertex can be in the aware \( (A) \) or unaware \( (U) \) state.
	When a susceptible vertex is aware, it will take precautions, effectively decreasing the probability of infection by a factor \( 0\leq\npi\leq1 \).

    To model the complex contagion of awareness, a linear threshold model is used.
	Here, a vertex adopts a state if the proportion of its neighbours in that state exceeds a given threshold \( \thres \)~\cite{Guo2015}.
	In our model, a convex combination governed by \( \comb \), referred to as the combination parameter, is used to compare this threshold with:
	\begin{linenomath}\begin{equation}
		\Ca = (1-\comb)\frac{\nbsA}{\degOpi} + \comb\frac{\nbsI}{\degEpi}. \label{eq:convex}
	\end{equation}\end{linenomath}
	Here, \( \nbsA \) and \( \nbsI \) represent the number of aware and infectious neighbours, and \( \degOpi \) and \( \degEpi \) the degrees of vertex \( i \) in the social and \physical~layer, respectively.
	The parameter \( \comb \) weights the importance of the neighbourhood status in the \physical\ layer and in the social layer (\Figs~\ref{subfig:opi_dynamics_0} to~\ref{subfig:opi_dynamics_10}).
	Varying \( \comb \) allows us to study the behaviour based on local risk perception (\( \comb=1 \)) and a pure threshold model (\( \comb=0 \)).
	In the former case, an individual only takes preventive actions based on the proportion of sick neighbours, i.e.\ it does not base its belief on that of others, but only on what it perceives in its direct neighbourhood.
	At the other side (\( \comb=1 \)), the awareness state of a vertex only depends on the proportion of aware neighbours, independently of the infection dynamics.
	That is, a vertex bases its awareness state on what it hears from others, and disregards its physical surroundings.
	Since \Eq~\eqref{eq:convex} is a convex combination, if neither of the two proportions exceeds \( \thres \) for a vertex \( i \), \( \Ca < \thres \) holds and the vertex will not become aware for any value of \( \comb \).

	At each time step, first the awareness dynamics are calculated, using the following rule,
	\begin{linenomath}\begin{subequations}
		\begin{align}
			\Pit{XY}{AY} &= H\left(\thres - \Ca\right), \label{eq:modelstart}\\
			\Pit{XY}{UY} &= 1 - \Pit{XY}{AY}, \label{eq:awarenessend}
		\end{align}
	\end{subequations}\end{linenomath}
	where \( X \) and  \( Y \) represent any possible awareness and disease state, respectively, and \( H \) is the Heaviside step function.
	This process is thus deterministic, since the transition probability is either 0 or 1.
	Next, the infections are simulated as a function of the vertex's awareness state and the infection state of its \physical~neighbours, according to the transition probabilities:
	\begin{linenomath}\begin{subequations}
		\begin{align}
			\Pit{US}{US} &= {(1-\Pinf)}^{\nbsI}, \\ \label{eq:epistart}
			\Pit{AS}{AS} &= {(1-\npi\Pinf)}^{\nbsI}, \\
			\Pit{XS}{XI} &= 1 - \Pit{XS}{XS}, \\
			\Pit{XI}{XI} &= 1 - \Prec, \\
			\Pit{XI}{XR} &= \Prec,\label{eq:modelend}
		\end{align}
	\end{subequations}\end{linenomath}
	where \( \nbsI \) is the number of infectious neighbours of vertex \( i \), \( \Prec \) the probability that an infectious vertex recovers, and \( \Pinf \) the probability that it infects a susceptible neighbour (\Fig~\ref{subfig:epi_dynamics}).

	As a model extension, we add trusted, static awareness sources to the model.
	These are virtual vertices on the social layer.
	There is an unaware, and an aware one, and both never change their awareness state.
	Every vertex is either connected to the aware one with a probability \( \trust \), or to the unaware one with probability \( 1-\trust \).
	In practice, this means that the convex combination \( \Ca \) (\Eq~\eqref{eq:convex}) changes to
	\begin{align}
		\Cua &= (1-\comb)\frac{\nbsA}{\degOpi + 1} + \comb\frac{\nbsI}{\degEpi}, \label{eq:convex-unaware}\\
	\intertext{for the vertices connected to the unaware, and}
		\Caa &= (1-\comb)\frac{\nbsA + 1}{\degOpi + 1} + \comb\frac{\nbsI}{\degEpi}, \label{eq:convex-aware}
	\end{align}
	for vertices connected to the aware institution.
	These static vertices are used to study the effect of external, trusted sources of information on the awareness dynamics in \Sec~\ref{sec:trust}.

	\subsection{Experiments}\label{sec:setup}
	The model of \Sec~\ref{sec:dynamics} is analysed with Monte Carlo simulations.
	Values of interest are the fractions  of aware \( A\t \) and infectious \( I\t \) vertices over time, the awareness at steady state \( A^\infty \) in function of \( \comb \) and \( \thres \), and the final fraction of recovered vertices \( R^\infty \) under influence of \( \Pinf \).
	A global transmission reduction factor \( \grf(\comb, \thres) \) allows to quantify the effectiveness of the protective behaviour induced by the awareness.
	For the calculation of this factor, we extract the infection state of every vertex at each time step, discarding the awareness.
	From this time series \( \Pinf^*(\comb, \thres) \) is inferred, which is the estimated transmission probability for an infection process on the \physical~layer without awareness.
	The global reduction factor, given by
	\begin{equation}
		\grf(\alpha, \thres)=\Pinf^*(\alpha, \thres)/\Pinf,\label{eq:reduction}
	\end{equation}
	then allows to quantify how much the awareness reduces the risk of infection on average over all vertices for a certain parameter combination \( (\comb, \thres) \).
	A measure of the intensity of the infection is the peak infection proportion \( PI(\alpha) \), which is the maximum fraction of infectious vertices over time,
	\begin{equation}
		PI(\alpha') = \max_t I\t|_{\alpha=\alpha'}\label{eq:PI}
	\end{equation}
	for a given awareness parameter \( \alpha' \).
	In order to study why a vertex becomes aware in \Sec~\ref{sec:beta}, we introduce
	\begin{subequations}\label{eq:nu}
		\begin{align}
			\nu_A^A\t = \frac{1}{NA\t}\mathlarger{\mathlarger{\sum}}_{i\in\mathcal{V}^A\t}\frac{l^A_i\tm}{l_i}, \\
			\nu_A^I\t = \frac{1}{NA\t}\mathlarger{\mathlarger{\sum}}_{i\in\mathcal{V}^A\t}\frac{k^I_i\tm}{k_i},
		\end{align}
  \end{subequations}
	where \( \mathcal{V}^A\t \) denotes the set of aware vertices at time step \( t \).
	In words, \( \nu_A^A \) (\( \nu_A^I \)) is the average fraction of aware (infectious) neighbours of vertices that become aware.

	The presented results are averaged over 5 different realisations of the network, and 25 distinct trajectories per network model, each starting from a different initial configuration and considered up to the point where no vertex is infectious and the awareness reaches a steady-state.
	First the single-layer model networks, i.e.~those where the \physical~and social layers coincide, are considered, then the multiplex ones.
	The networks consist of 4096 vertices.
	Unless otherwise mentioned, the model parameters are \( \Pinf=0.2 \), \( \Prec=0.15 \), \( \comb=0.5 \), \( \thres=0.45 \) and \( \npi=0 \), i.e.\ awareness completely inhibits infections.
	As for the initial condition, four random vertices in the \physical~layer are chosen to be infectious, and each vertex in the social layer is aware with probability \( 0.5 \), unless otherwise mentioned.

	\section{Results and discussion}\label{sec:results}

	\subsection{Single-layer networks}\label{sec:single}

	\subsubsection{Dynamical evolution}\label{sec:single_dynamical}

\begin{figure*}[t]
		\centering
		\graphsquare{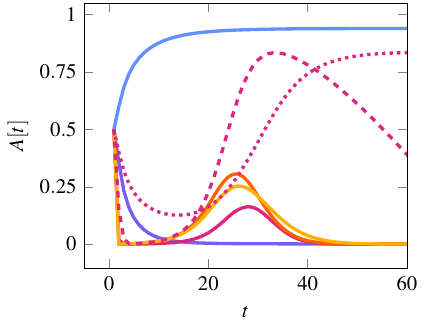}{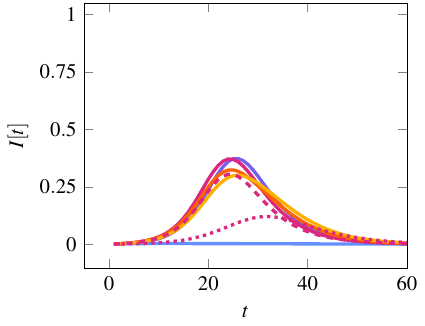}{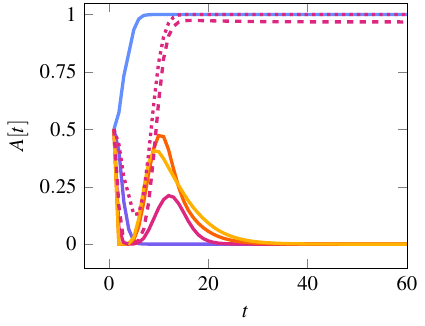}{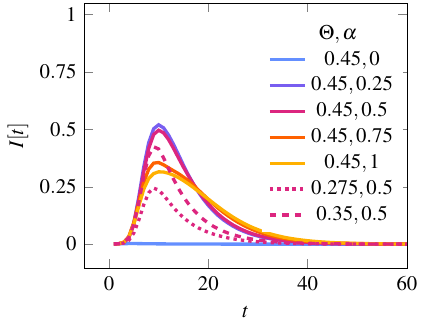}
		\caption{%
			Temporal evolution of the fractions of (\subref{subfig:t_sw_A}, \subref{subfig:t_sf_A}) aware and (\subref{subfig:t_sw_I}, \subref{subfig:t_sf_I}) infectious vertices on (\subref{subfig:t_sw_A}, \subref{subfig:t_sw_I}) \SW{}~and (\subref{subfig:t_sf_A}, \subref{subfig:t_sf_I}) \SF{}~networks for several values of the combination parameter \( \comb \) (by colour) and threshold \( \thres \) (by line type).
		}%
		\label{fig:temporal_single}
	\end{figure*}

	\Fig~\ref{fig:temporal_single} shows the evolution of the fraction of aware \( (A\t) \) and infectious \( (I\t) \) vertices for the case where \( \thres=0.45 \) (full line) for different values of \( \comb \), on single-layer small-world (SW), and single-layer scale-free (SF) networks.
	There is a discrepancy in timescale between the awareness and infection dynamics.
	If \( \comb=0 \) an awareness cascade occurs.
	As the aware state rapidly expands to the entire network, it completely inhibits the infection from spreading.
	In contrast, for the considered values where \( \comb \neq 0 \) every vertex becomes unaware, and this happens more quickly as \( \comb \) is larger.
	Then, the number of infections only becomes significant when the initial awareness has already died out.
	There is no interaction between the awareness due to the initial condition and due to the infection prevalence.
	For \( \comb=0.25 \), there is also no awareness generated by the number of infections, and there are thus no awareness effects on the course of the infections.

	As in the case of an infection spreading without awareness response, the clustering of the \SW{}~networks has an inhibitory effect on the infections, in comparison with the effect of the heterogeneous degree distribution of the \SF{}~networks~\cite{Nunner2022, PastorSatorras2001}.
	For every combination of \( \comb \) and \( \thres \), the peak of infections is lower and occurs at a later time step in the \SW~networks than in the \SF~ones.
	Especially at the initial time steps, the rapid increase of \( I\t \) in \Fig~\ref{subfig:t_sf_I} contrasts with the more smooth onset of the infections in \Fig~\ref{subfig:t_sw_I}.

	Giving more importance to local risk perception increases the inhibitory effect of the awareness of the infections.
	The monotonously decreasing infection peaks for growing \( \comb>0 \) (\Figs~\ref{subfig:t_sw_I} and~\ref{subfig:t_sf_I}) show this.
	The maximum infection proportion (\Eq~\eqref{eq:PI}) decreases from \( PI(0.25)=0.375\pm0.007 \) to \( PI(1)=0.300\pm0.005 \) in the case of the \SW{}~model, and from \( PI(0.25)=0.527\pm0.012 \) to \( PI(1)=0.317\pm0.001 \) on the \SF{}~networks.
	Here, and in what follows, it is the mean \( \pm \) the standard deviation over the different network realisations, that is reported.
	Local risk perception decreases the peak infectious proportion more in a \SF{} than a \SW{} network, in contrast to an awareness cascade that has the same inhibiting effects on both network models.
	When \( \comb=1 \), there is only a small difference in peak infection proportion between the two network models, as opposed to when there are no awareness effects.
	The time to peak infection \( (\text{argmax}_t~I\t) \) is not affected as much, happening around time step 25 for the \SW{}~network and time step 10 for the \SF{}~network, irrespective of the value of \( \comb \).

	For the trajectory of the number of aware vertices, there is a more complex process in play for the values of \( \comb \) equal to 0.5, 0.75 and 1.
	Initially, the peak proportion of aware vertices rises with increasing \( \comb \), however, this relationship is non-monotonous.
	\Figs~\ref{subfig:t_sw_A} and~\ref{subfig:t_sf_A} show that for the maximum value of \( \comb \), there is less awareness than for \( \comb=0.75 \).
	This is the result of a direct and indirect effect.
	First, the higher \( \comb \), the larger the influence of the infections on the awareness generation, so the closer the trajectory of the awareness density will be to that of the infection prevalence.
	Secondly, this extra awareness lowers the incidence of infectious vertices, such that the awareness is indirectly lowered as well.
	The time step of the onset of the awareness wave is similar to their infection proportion counterpart for the values of \( \comb \) equal to 0.5, 0.75 and 1.
	This is because the initial awareness fades, and the subsequent awareness is thus generated by infectious neighbours.
	Lower values of \( \comb \) cause a delay in the peak time of the aware proportion compared to the peak time of the infectious proportion.
	Like the simple contagion, the complex contagion spreads or dies out, if \( \comb=0 \) or \( \comb=0.25 \) respectively, more rapidly on the  \SF{}~network than on the \SW{}~network.

	\subsubsection{Steady-state}\label{sec:single_steady}

	\begin{figure*}[t]
		\heatmapsquare{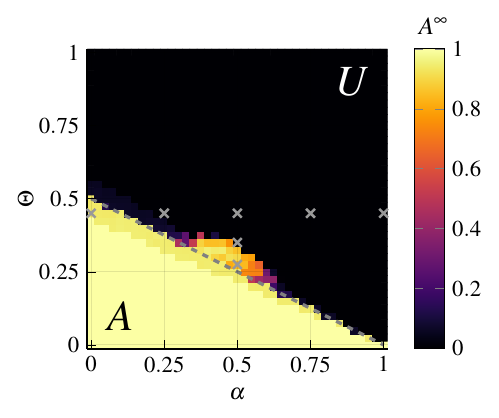}{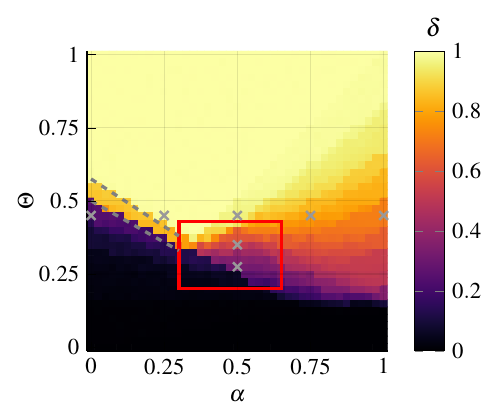}{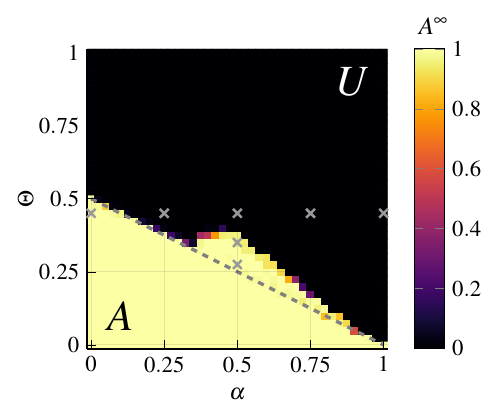}{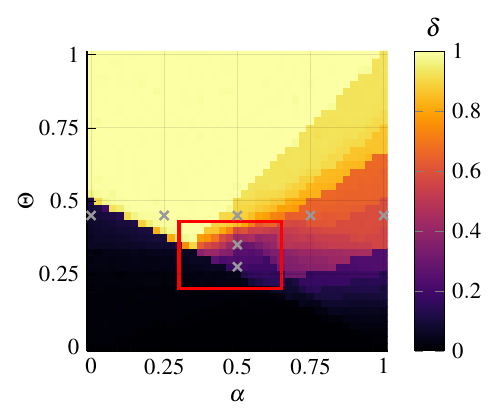}
		\caption{%
			Final fractions of (\subref{subfig:sw_A}, \subref{subfig:sf_A}) aware vertices and (\subref{subfig:sw_bs}, \subref{subfig:sf_bs}) global reduction factor in the \( (\comb,\thres) \)-parameter space on (\subref{subfig:sw_A}, \subref{subfig:sw_bs}) single-layer \SW{}~and (\subref{subfig:sf_A}, \subref{subfig:sf_bs}) \SF{}~networks.
			In the left panels the aware (unaware) region is denoted with \( A \) (\( U \)), the dashed grey line represents the \( \thres=(1-\comb)/2 \) straight.
			The red rectangle indicates the region of complex behaviour and the crosses indicate the parameter combination for which the dynamical evolution is given in \Fig~\ref{fig:temporal_single}.
		}%
		\label{fig:single_steady_heatmaps}
	\end{figure*}

	\Fig~\ref{fig:single_steady_heatmaps} shows the fractions of aware vertices at steady-state and the global reduction factor \( \grf \) in function of the combination parameter \( \comb \) and threshold \( \thres \).
	The awareness state (\Figs~\ref{subfig:sw_A} and~\ref{subfig:sf_A}) is homogeneous over the networks for most parameter combinations.
	This results in one region in which every vertex is aware, another one in which none is.
	For the lower half plane below the straight line \( \thres=(1-\comb)/2 \) (\Figs~\ref{subfig:sw_bs} and~\ref{subfig:sf_bs}), it is the initial awareness that spreads rapidly to the complete network.
	The dynamical evolution for \( \comb=0 \) (\Figs~\ref{subfig:t_sw_A},~\ref{subfig:t_sf_A}), and the black area (\( \grf \) = 0) in the parameter spaces (\Figs~\ref{subfig:sw_bs},~\ref{subfig:sf_bs}) show this.
	These results indicate that the awareness reduces the probability of an infection event to zero.
	Because initially only a negligible fraction of vertices is infectious, they do not have much influence on the awareness propagation and the second term of \Eq~\eqref{eq:convex} may be ignored, resulting in the straight line \( \thres=(1-\comb)/2 \) (grey dashed line \Fig~\ref{fig:single_steady_heatmaps}).
	In other words, when the initial awareness produces a cascade it holds that \( \thres/(1-\comb) \leq 1/2 \).
	Here, \( \thres/(1-\comb) \) is the threshold rescaled by the awareness contribution weight.
	If this condition is not met, awareness dies out quickly (\Figs~\ref{subfig:t_sw_A} and~\ref{subfig:t_sf_A}).

	An additional region in the parameter space allows for an aware stable state.
	Situated around \( \comb=0.5 \), \( \thres<0.5 \), it is more pronounced for the \SF{}~network (\Fig~\ref{subfig:sf_A}) than for the \SW{}~network (\Fig~\ref{subfig:sw_A}).
	This follows from the difference in the effect of topology on spreading processes on these networks.
	Clustered networks such as \SW{}~models inhibit the spreading of biological infections, while \SF{}~networks facilitate this process~\cite{Nunner2022, PastorSatorras2001}.
	The latter allows the number of concurrent infectious vertices to rise to a level that generates an aware seed able to cause a cascade through the second term of \Eq~\eqref{eq:convex}~\cite{Gleeson2007}.
	In the case of the \SW{}~network, the cascades are less pronounced.
	The area of the region in the parameter space in which they occur, as well as their mean size is smaller than in the case of the \SF{}~network.
	This is due to the higher level of clustering in the \SW{}~network, which lowers the number of concurrently infectious vertices and reduces the cascade size~\cite{Nunner2022, Hackett2011}.
	To better illustrate this phenomenon, \Fig~\ref{fig:temporal_single} also shows the fractions of infectious and aware vertices in both networks for \( \comb=0.5 \) and \( \thres=0.275 \), \( \thres=0.35 \).
	For both threshold values,  an awareness cascade occurs in the \SF{}~network (\Fig~\ref{subfig:t_sf_A}) due to the rising number of infections.
	The situation of the \SW{}~network (\Fig~\ref{subfig:t_sw_A}) is more complex.
	For \( \thres=0.275 \), a cascade occurs, but it does not reach the complete network, resulting in a polarised awareness steady-state with both aware \( (0.834\pm0.032) \) and unaware \( (0.166\pm0.032) \) vertices.
	For \( \thres=0.35 \), the peak value of the aware proportion is high as well \( (0.831\pm0.014) \).
	That level, however, cannot be sustained when the number of infectious vertices diminishes, eventually resulting in a totally unaware state.
	For these complex phenomena to occur, the value of the combination parameter \( \comb \) ought to be in an intermediate range, since the fraction of infections needs to have enough weight such that awareness can be generated from infectious neighbours.
	Besides, there also needs to be a contribution from the social layer in \Eq~\eqref{eq:convex}, such that the awareness can spread over the entire network to vertices that do not have infectious neighbours.
    This effect is examined more closely in \Sec~\ref{sec:beta}.

	\Figs~\ref{subfig:sw_bs} and~\ref{subfig:sf_bs} show the global reduction factor \( \grf \) (\Eq~\eqref{eq:reduction}) to assess the influence of awareness on the infection process.
	For low values of the combination parameter \( \comb \), meaning awareness mainly comes from aware neighbours, the all-or-nothing results of awareness cascade effects are noticeable again.
	For the \SF{}~networks (\Fig~\ref{subfig:sf_bs}) this effect is more pronounced than for the \SW{}~networks (\Fig~\ref{subfig:sw_bs}), for which there is a limited region around the straight line \( \thres=(1-\comb)/2 \) straight with intermediate proportions of aware vertices in steady state, as bounded by the grey dashed lines.
	This can again be attributed to the ability of clustered networks to support smaller cascade sizes.
	For \( \comb=1 \), meaning that a vertex becomes aware and takes precautions due to infectious vertices in its neighbourhood, the changes are gradual for both networks, that is, from no effect if \( \thres=1 \) to no infection events happening if \( \thres=0 \).
	This is because on the former end precautions are only taken when all neighbours are infectious simultaneously, while on the latter end a vertex is able to perfectly protect itself as soon as it may become infectious.
	In contrast to the awareness cascades at \( \comb=0 \), the switch does not happen at a single value of the threshold, with \( \grf \) taking intermediate values between 0 and 1 for increasing \( \thres \).

\begin{figure}[t]
		\centering
		\graphrow{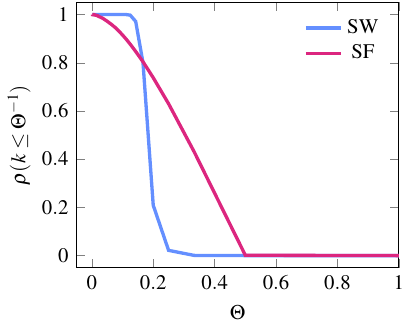}{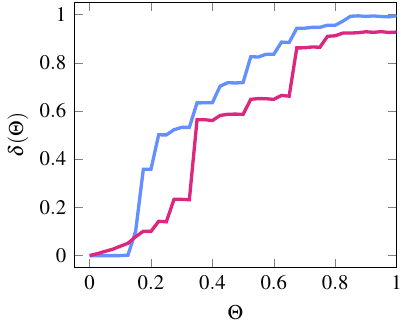}
		\caption{%
			(\subref{subfig:rho}) The proportion of vertices with degree lower than or equal to \( 1/\thres \), \( \rho(k\leq\thres^{-1}) \) and (\subref{subfig:delta}) the global reduction factor \( \grf \) at \( \comb=1 \) (dashed) for \SW~(blue) and \SF~(purple) network models.
		}%
		\label{fig:f_theta}
	\end{figure}

	For low values of \( \thres \) and \( \comb=1 \) the effect of the different network topologies is apparent.
	In this case of local risk perception, it is impossible for a vertex with degree lower than or equal to \( 1/\thres \) to become infectious, since it will take perfect precautions as soon as one neighbour is infectious.
	This impacts \( \grf \) depending on the network topology (\Fig~\ref{fig:f_theta}).
	Here, a slice of \Figs~\ref{subfig:sw_bs} and~\ref{subfig:sf_bs} is shown at \( \comb=1 \), along with the proportion of vertices in the \SW{}~and \SF{}~networks for which the degree is lower or equal to the inverse of the threshold, \( \rho(k\leq\thres^{-1}) \).
	Because of the relatively narrow degree distribution centred around \( k=6 \) of the \SW{}~network, the change between almost none (\( \rho(k\leq\thres^{-1})\approx1 \)) and almost all of the vertices (\( \rho(k\leq\thres^{-1})\approx0 \)) are infectable, happens over a narrow range of \( \thres \)-values.
	For \( \thres<0.125 \), i.e.\ lower than this range, \( \grf \) is thus equal to zero, but it grows rapidly as soon as \( \rho(k\leq\thres^{-1}) \) decreases.
	In the case of the \SF{}~network, this is different.
	The proportion of vertices that is not at risk of infection, decreases less abruptly over the interval \( \thres\in[0,0.5] \).
	As a result, the awareness has an effect for low values of \( \thres \) and the global reduction factor \( \grf \) takes a value larger than zero, i.e.\ the effects of awareness are in this case more stringent on the \SW{}~networks than on the \SF{}~ones.
	However, as \( \thres \) becomes greater than 0.15, the value of \( \grf \) for the \SW{}~networks overtakes that of the \SF{}~networks.
	This is another indication that the local risk perception has a larger effect on \SF{}~networks, as shown before in \Figs~\ref{subfig:t_sw_I} and~\ref{subfig:t_sf_I}.

	The global reduction factor \( \grf \) does not vary smoothly with respect to \( \thres \) for \( \comb=1 \), but changes at values that are network dependent.
	Similar to the lower end of the \( \comb \)-range, the values at which these discontinuities happen, scale linearly with \( \comb \).
	So, it can thus be reasoned that now the first term in \Eq~\eqref{eq:convex} may be ignored.
	This means that for high values of \( \comb \) the pure awareness spread between neighbours does not play a significant role.
	Between the regions where each of the terms of \Eq~\eqref{eq:convex} is dominant, we find non-trivial, complex behaviour.
	This is again the region near \( \comb\approx0.5 \), \( \thres<0.5 \), that has already been discussed in the previous paragraph.
	This region is enclosed by the red rectangle for both networks in \Figs~\ref{subfig:sw_bs} and~\ref{subfig:sf_bs}.
	Here, both the awareness generation by infectious neighbours and the awareness spreading by aware neighbours have an influence on the infection process.

	\begin{figure*}[t]
		\centering
		\heatmaprow{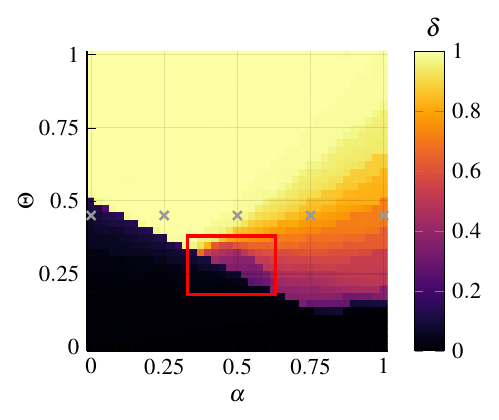}{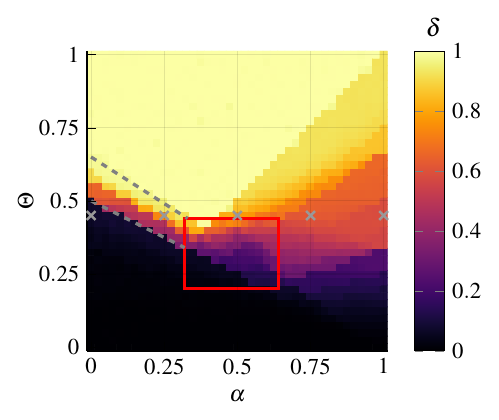}
		\caption{%
			The global reduction factor \( \grf \) on multiplex (\subref{subfig:swsf_bs}) \SWSF~and (\subref{subfig:sfsw_bs}) \SFSW~networks.
			The region with non-trivial behaviour is enclosed by the red rectangle, and the crosses indicate the parameter combination for which the dynamical evolution is given in Fig~\ref{fig:temporal_multi}.
		}%
		\label{fig:beta_star}
	\end{figure*}

	\subsection{Multiplex networks}\label{sec:multiplex_results}
	\subsubsection{With initial awareness}
	The multiplex networks are more realistic, because the \physical\ and social connections do not coincide\ \cite{Brodka2020}.
	We consider combinations of the model networks in \Secs~\ref{sec:single_dynamical} and~\ref{sec:single_steady}, in particular, \SWSF\ and \SFSW\ multiplex networks, where the network models refer to the \physical\ and social layers, respectively.
	For values of \( \comb \) close to 1, the global reduction factor \( \grf \) on the \SWSF~(\Fig~\ref{subfig:swsf_bs}) and \SFSW~(\Fig~\ref{subfig:sfsw_bs}) network shows a similar dependence on \( \thres \) and \( \comb \) as on their respective \physical~layers (\Figs~\ref{subfig:sw_bs} and~\ref{subfig:sf_bs}).
	For such values of \( \comb \), the \physical~layer is the dominant one.
	At the lower end of the \( \comb \)-range, the sudden transition between no awareness and a fully aware population is discernible again.
	There are, however, quantitative differences between the single- and double-layer topologies.
	In the case of the \SWSF~network, the boundary between the two regions is more pronounced than in the case of the \SF{}~networks.
	In contrast, the range of \( \thres \)-values with intermediate steady-state awareness proportions at \( \comb=0 \) is the largest for the \SFSW~networks (grey dashed lines, \Fig~\ref{subfig:sfsw_bs}).
	This is due to the difference in time scales, not only between the two processes, but also between the two network models.
	On the same network, awareness spreads faster than the infection, while the same process spreads faster on the \SF{}~network than on the \SW{}~network (\Fig~\ref{fig:temporal_single}).
	In the case of the \SWSF~topology we have a combination of a rapid process (awareness) on a facilitating layer (\SF{}), and a slower spreading process (infections) on a inhibiting layer (\SW{}).
	Consequently, the difference in timescales discussed in \Sec~\ref{sec:single_dynamical} is amplified, resulting in a sharper boundary between the fully aware and unaware states.

   	\begin{figure*}[t]
   		\centering
   		\graphsquare{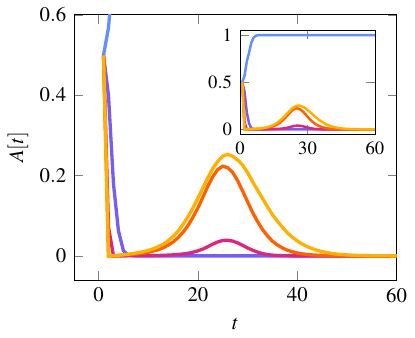}{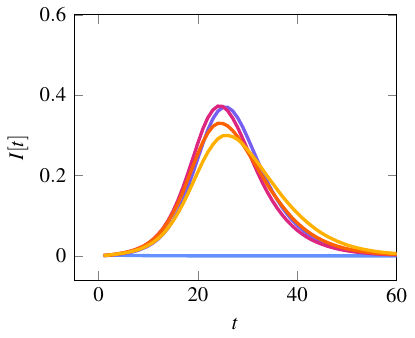}{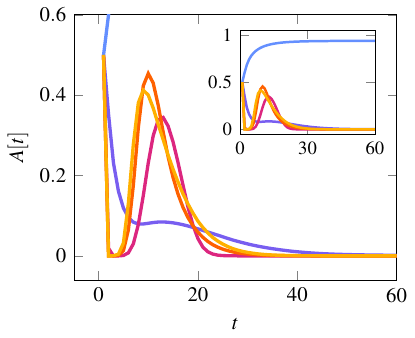}{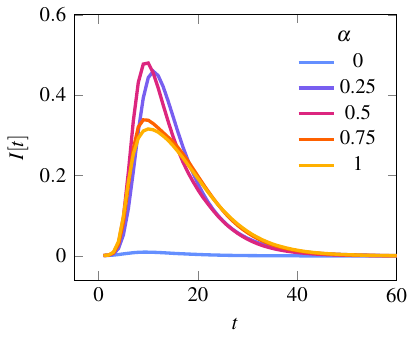}
   		\caption{%
   			Temporal evolution of the fractions of aware (left) and infectious (right) vertices on \SWSF~(top) and \SFSW~(bottom) multiplex networks, for several values of the combination parameter \( \comb \) and \( \thres=0.45 \).
   			The insets show the full range of the aware proportions.
   		}%
   		\label{fig:temporal_multi}
   	\end{figure*}

	For \SFSW~networks the opposite holds, allowing the infections and awareness to interact over longer periods.
	In \Fig~\ref{subfig:t_sfsw_A}, in contrast to the single-layer (\Figs~\ref{subfig:t_sw_A} and~\ref{subfig:t_sf_A}) and \SWSF\ (\Fig~\ref{subfig:t_swsf_A}) networks, the number of infections rises faster than the initial awareness can die out for \( \comb=0.25 \), and there are awareness effects due to both the initial awareness and rising infections.
	The awareness inhibits infection events (\Fig~\ref{subfig:sfsw_bs}), and the maximum fraction of infectious vertices is higher for \( \comb=0.5 \) than for \( \comb=0.25 \), in contrast to the other networks.
	However, as the infections lessen, the awareness cannot sustain itself and subsides as well.
	For other values of \( \comb \), the trajectories of the multiplex networks are similar to those of their \physical~layer.
	This follows from the extinction of the initial awareness, after which the infections on the \physical~layer (second term of \Eq~\eqref{eq:convex}), influence the awareness state of a vertex.

	\subsubsection{Influence of transmission probability}\label{sec:beta}

   	\begin{figure*}[t]
   		\centering
      \graphrow{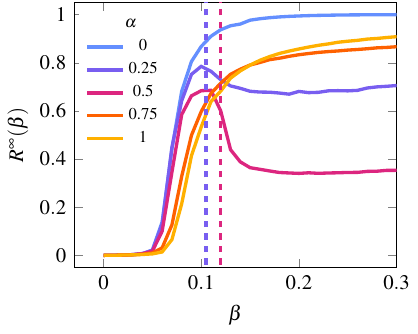}{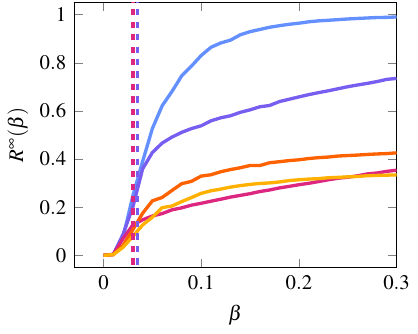}
   		\caption{%
				Dependence of the final fraction of recovered vertices in (\subref{subfig:l_swsf_R}) \SWSF~and (\subref{subfig:l_sfsw_R}) \SFSW~multiplex networks on the transmission probability for various values of \( \comb \) and \( \thres=0.25 \).
				If an awareness cascade occurs for a value of \( \comb \), the corresponding value of \( \Pinf \) is indicated by a dashed line in the same colour.
   		}%
   		\label{fig:R_beta}
   	\end{figure*}

	To study the emergence of awareness cascades due to infections, we measure the fraction of recovered vertices versus the transmission probability (\Fig~\ref{fig:R_beta}).
  In this experiment, no vertices are initially aware.
	Thus if \( \thres/\comb>1 \), there will be no awareness in this set-up.
	To have awareness effects for most values of \( \comb \), the threshold is chosen to be \( \thres=0.25 \).
	For \( \comb=0 \) there is no awareness and the proportion of recovered vertices behaves as for a typical SIR process.
	For \( \Pinf \) lower than the epidemic threshold, the proportion of recovered vertices is negligible~\cite{Wang2016b}.
	Once \( \Pinf \) exceeds this threshold, \( R^\infty \) rises.
	This epidemic threshold is lower for the \SF{}~model than for the \SW{}~structure.
	In both networks, there is no awareness cascade for \( \comb=0.75 \) and \( \comb=1 \).
	While the \SF{}~topology accelerates the infection process, the effect of local-risk response is larger as well.
	For \( \Pinf=0.3 \) nearly the entire population in both networks is recovered, if \( \comb=0 \).
	However for pure local-risk response (\( \comb=1 \)), this reduces to proportions of \( 0.908\pm 0.003 \) for the \SWSF, and \( 0.335\pm0.008 \) for the \SFSW~network.
	The larger effect in the latter case is a consequence of the difference in degree distribution (see \Sec~\ref{sec:single_steady}).
	The \SF~network contains many vertices of degree equal to three, which will take precautions with only one infectious neighbour, even for \( \comb=0.75 \).

	Both in \SFSW~(\Fig~\ref{subfig:l_swsf_R}) and \SWSF~(\Fig~\ref{subfig:l_sfsw_R}) networks, awareness cascades occur for values of \( \comb=0.25 \) and \( \comb=0.5 \).
	For this to happen, vertices have to first become aware because of their infectious \physical~neighbours, and then have to spread the awareness to their social neighbours.
	Under the considered conditions, both these processes happen more easily on the \SF~network (\Sec~\ref{sec:single_dynamical}).
	For the \SFSW~network (\Fig~\ref{subfig:l_sfsw_R}), the awareness cascades occur for lower \( \Pinf \) than for the \SWSF~network (\Fig~\ref{subfig:l_swsf_R}), since the initial seed of infectious vertices forms more easily.
	However, when awareness cascades occur, the effects are more drastic for the \SFSW~network, even resulting in a decrease in infection reach for growing \( \Pinf \) near the value that the cascade occurs.

	\begin{figure}[t]
   		\centering
   		\graphrow{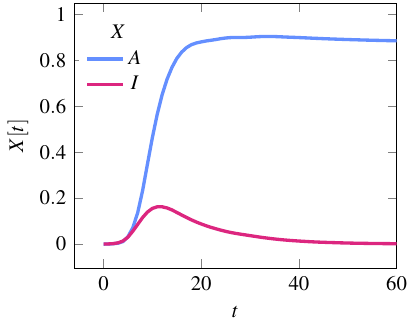}{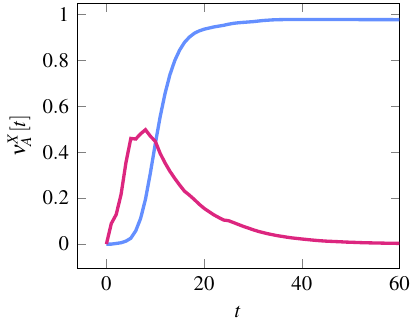}
   		\caption{%
           (\subref{subfig:t_sfsw_why_A}) \( A\t \), \( I\t \), (\subref{subfig:t_sfsw_why_A_nu}) \( \nu_A^A\t \) and \( \nu_A^I\t \) on a \SFSW~multiplex network for \( \comb=0.5 \) and \( \thres=0.25 \).
        }%
   		\label{fig:why_as}
   	\end{figure}

	\Fig~\ref{fig:why_as} shows the two-step process of the emergence of an awareness cascade in a \SFSW~network with no initial awareness.
	The global fractions of aware \( A\t \) and infectious \( I\t \) vertices are given over time, along with the mean fractions of infectious and aware neighbours of aware vertices, \( \nu_A^I\t \) and \( \nu_A^A\t \) (\Eq~\eqref{eq:nu}).
	Initially, vertices become aware because of infectious neighbours.
	These infectious vertices are localised in the \physical~layer, forming a cluster, shown by the value of \( \nu_A^I\t \) which is larger than \( I\t \), the mean over the complete network that is close to zero for early time steps.
	Later, around \( t=8 \), the awareness starts inhibiting the disease, as well as aiding the remaining infectious vertices in making their neighbours aware.
	Since the connections of the social and \physical~layers are not correlated, the aware neighbours are distributed randomly, as reflected by the coincident trajectories of \( \nu_A^A\t \) and \( A\t \).
	This changes around time step 18 when many of the aware vertices only have aware neighbours and the reduction in infections also inhibit the becoming aware of the few final unaware vertices.
	In other words, whether the spread of a biological infection is inhibited by awareness, as might have been in the case of SARS~\cite{Durham2011, Shi2019}, depends on factors relating to how the infection, and awareness spreads, but also on the differences between \physical\ and social connections.

	\subsubsection{Influence of trusted sources}\label{sec:trust}

	\begin{figure*}[t]
		\centering
			\graphrow{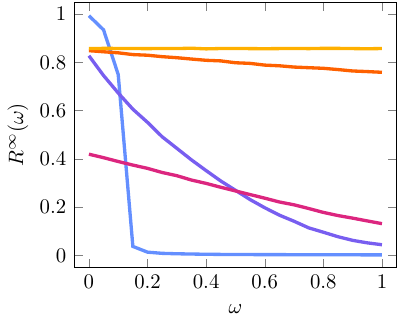}{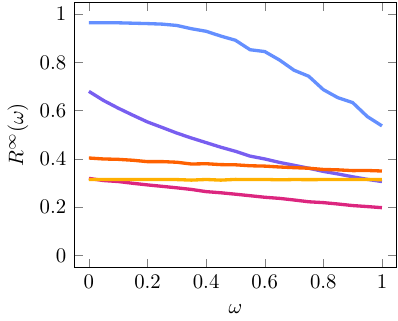}
		\caption{%
			Dependence of the final fraction of recovered vertices in (a) SW-SF and (b) SF-SW multiplex networks on the fraction of vertices connected to the aware trusted source \( \trust \) for various values of \( \comb \) and \( \thres=0.25 \).
		}%
		\label{fig:R_omega}
	\end{figure*}

	In the final experiment, two trusted awareness sources are added to the social layer, one aware and one unaware.
	A proportion of \( \trust \) vertices is connected to the aware one, and of \( 1-\trust \) to the unaware one.
	The final fraction of recovered vertices is measured as a function of \( \trust \) in \Fig~\ref{fig:R_omega}.
	Like in \Sec~\ref{sec:beta}, the results are shown for various values of \( \comb \), there is no initial awareness, and the threshold is set to \( \thres=0.25 \).
	Interestingly, for all values of \( \trust \) and both networks, an awareness cascade occurs for \( \comb = 0.25 \) and \( \comb = 0.5 \), but not for \( \comb=0.75 \) and \( \comb = 1 \).
	For \( \comb = 0 \),	an awareness cascade happens if \( \trust > 0.15 \) the \SWSF~network, while for the \SFSW~network partial cascades occur from \( \trust = 0 \) that increase in size until they reach the complete network at \( \trust = 1 \).
	For both networks, the fraction of recovered vertices decreases with increasing \( \trust \).
	The rate at which this happens, decreases with increasing \( \comb \).
	\Eqs~\eqref{eq:convex-unaware}~and~\eqref{eq:convex-aware} offer an explanation.
	The importance of the first term in each diminishes with increasing \( \comb \), meaning that the (un)awareness due to the trusted source becomes less important.
	For \( \comb = 1 \) there are no effects and \( R^\infty(\trust) \) is constant.

	The rate of decrease of \( R^\infty(\trust) \) is larger for the \SWSF~network than for the \SFSW~network.
	This is due to the large number of low-degree vertices in the social layer of the \SWSF~network, which, when connected to the aware trusted source, will become aware more easily than high-degree vertices.
	Hence, the optimal value (in terms of minimal \( R^\infty \)) of \( \comb \), the importance of the social layer, depends both on the network topology and on the value of \( \trust \).
	For low values of \( \trust \), it is better to have a high \( \comb \), while high values of \( \trust \) benefit more from a low \( \comb \).
	For intermediate values of \( \trust \), low values of \( \comb \) are preferred for the \SWSF~network, but not for the \SFSW~network.
	In case of the latter, the optimal value of \( \comb \) is around 0.5, independent of \( \trust \), implying that both layers are always important for the curbing of the infections.
	For the optimal spread of awareness, it is thus helpful that most vertices base their state mainly on a trusted aware source.
	However, if that is not possible, because of a large presence of an unaware trusted source or other social contacts, it is best to take the infectious neighbours into account as well.

	\subsection{Differences with real-life epidemics}\label{sec:reallife}

	A stringent assumption of the model is the homogeneous nature of the thresholds, both over the vertices as over time.
	Realistically, one would expect that for real-world diseases not everyone requires the same level of infections or awareness in their neighbourhood to start taking precautions~\cite{Meijere2023}.
	Also past experiences may influence the tendency of an individual to become aware.
	Previous infections may cause an individual to be more inclined to take precautions.
	In contrast, the decrease in attention for the disease over time may increase the adoption threshold~\cite{Gozzi2020}.
	These factors are not taken into account in the current model, but could be included in future work as a time-dependent \( \thres \), or additional terms in \Eq~\eqref{eq:convex}.
	Similarly, the model neglects the temporary nature of interactions~\cite{Bansal2010, Rocha2013}.
	Edges may appear and disappear over time, as the contacts of individuals change.
	This can happen both naturally, or under influence of the infection or awareness states~\cite{Nunner2022}.
	An individual that is infectious, may reduce its face-to-face contacts as it does not feel well, as such lowering effective infection transmission, but at the same time remain well-connected on the social layer.
	On the latter, vertices may change their contacts to be part of a more like-minded neighbourhood, introducing clusters of aware vertices in the network topology.
	These in turn influence the spread of the infection, with the potential of making the population more susceptible~\cite{Salathe2008}.

	The above additions would reduce the gap between our in-silico approach and real-world diseases, and could be included in future work.
	However, in order to calibrate such an advanced model, one needs access to data to construct the \physical\ and social networks, and probe beliefs, behaviour and disease state.
	The high level of heterogeneity in such model requires a large sample size of individuals in order to sufficiently, and the time-dependence requires longitudinal data.
	Acquiring this data is a challenging task.

	\section{Conclusion}\label{sec:conclusion}
	We developed and studied an infection-awareness model on multiplex networks.
	The probability of an infection event depends not only on the disease, but also on the awareness state of vertices, which in turn depends on whether a convex combination of the proportion of its aware and infectious neighbours exceeds a threshold.

	With pure social awareness spreading, depending on the threshold, either the awareness rapidly vanishes and the infections spread without interference, or the awareness spreads across the entire network and the infections stop after a few iterations.
	In contrast, in the case of risk perception, the effect of precautions is gradual.
  Here, the \SF{}~network facilitates the spread of a disease compared to the \SW{}~network, but the effects of the local risk perception also reduce the number of infections more on the former network model than on the latter.
	For intermediate values of the convex combination parameter \( \comb \), there is nontrivial behaviour, where the awareness is generated by the infections, and spreads further, even with the possibility of causing an awareness cascade.
	The \SF{}~network facilitates these events with respect to a \SW{}~network, because of a higher fraction of concurrently infectious vertices.

	Multiplex networks allow more realistic and complex dynamics.
	Awareness due to infections can interact with awareness in the initial condition.
	This is the case when fast dynamics (awareness) on an inhibiting topology (\SW) are combined with slower dynamics (infections) on a facilitating topology (\SF).
  Also without initial awareness, cascades can occur, but the circumstances and effects depend both on the topology of the \physical~as that of the social layer.
	The former must allow for an aware seed to come into existence and the latter must be able to spread the awareness to the rest of the network.
	While for both the \SWSF~and \SFSW~networks awareness cascades occur, the effect is different.
	For a \SF~\physical~layer, awareness cascades occur for lower values of the transmission probability, since a sufficiently large seed of infectious vertices forms more likely.
	However, the effect is less drastic, because the \SW~social layer slows down the spread of awareness.
	In contrast, on the \SWSF~topology, cascades occur for higher values of \( \Pinf \), but the effects are more pronounced, resulting locally in a decrease of the reach of the infections for growing \( \Pinf \).
	Hence, the effect infection-inhibiting awareness cascades does not only depend on the dynamical properties of the infection and awareness spreading, but also on both the topology of the \physical~and social layers.
	The effects of the latter are corroborated by the inclusion of trusted awareness sources, which have a more pronounced effect on \SF~social layers.
	In this case, an infection can be stopped purely due to trusted sources, tough in general it is best to also take the infectious neighbours into account.

	This study offers descriptions of the conditions under which awareness cascades occur, in a coupled epidemiological-awareness setting.
	However, many real-life aspects are still neglected, e.g.\ the temporary nature of interactions, and the existence of purely virtual vertices such as social media bots~\cite{Brodka2020, Rocha2013}.
	The networks presented herein are static and artificial.
	A threshold is not the only way to model awareness spreading, and leaves out many aspects of human behaviour~\cite{Verelst2016}.
	Also, every edge and vertex in the networks abide by the same rules and parameters, disregarding individual heterogeneity~\cite{Guilbeault2018, Meijere2023}.
	These factors inhibit the model from producing epidemiological forecasts.

	\section*{Acknowledgments}
	\subsection*{Funding}
	This study was supported by the Research Foundation --- Flanders (FWO) [G0G0122N].
	Computational infrastructure was provided by the Flemish Supercomputer Center (VSC).
	The two organisations did not in any way influence the design, execution or results of this work.

	\subsection*{Data statement}
	All data in this report has been generated by simulations.
	The code to redo to evaluate the model is available at the public Github repository \url{https://github.com/TimVWese/SIRLT.jl}.

	\subsection*{Conflict of interest}
	The authors report no conflict of interest.

\bibliographystyle{unsrtnat}
\bibliography{literatuur}
\end{document}